\documentclass[conference]{IEEEtran}
\IEEEoverridecommandlockouts
\usepackage{amsmath,amssymb,amsfonts}
\usepackage{algorithmic}
\usepackage{graphicx}
\usepackage{textcomp}
\usepackage{xcolor}
\usepackage{colortbl}
\usepackage{multirow}
\usepackage{tabularx}

\usepackage{wrapfig}

\usepackage{mathtools}
\usepackage{adjustbox}
\usepackage{subfigure}

\def\BibTeX{{\rm B\kern-.05em{\sc i\kern-.025em b}\kern-.08em
    T\kern-.1667em\lower.7ex\hbox{E}\kern-.125emX}}
\usepackage{array}
\newcolumntype{L}[1]{>{\raggedright\let\newline\\\arraybackslash\hspace{0pt}}m{#1}}
\newcolumntype{C}[1]{>{\centering\let\newline\\\arraybackslash\hspace{0pt}}m{#1}}
\newcolumntype{R}[1]{>{\raggedleft\let\newline\\\arraybackslash\hspace{0pt}}m{#1}}

\newcommand{\ie}{{\em i.e., }}
\newcommand{\eg}{{\em e.g., }}

\newcommand{\mycomment}[1]{}

\begin{document}

%\title{Assessing Metaverse Virtual Reality User Experience under Degraded Network Conditions}
%\title{Relating Metaverse VR QoE to Network QoS in Public and Private Virtual Spaces}
%\title{Assessing the Impact of Network Degradation on Metaverse VR User Experience in Public and Private Virtual Spaces}
\title{Assessing the Impact of Network Quality-of-Service on Metaverse Virtual Reality User Experience\thanks{This work is supported by funding from Canopus Networks Pty Ltd.}}

\author{
 \IEEEauthorblockN{Rahul Dev Tripathi}
 \IEEEauthorblockA{%\textit{dept. name of organization (of Aff.)} \\
 	\textit{University of New South Wales}\\Sydney, NSW, Australia\\
 	r.tripathi@student.unsw.edu.au}
 \and
 \IEEEauthorblockN{Minzhao Lyu}
 \IEEEauthorblockA{%\textit{dept. name of organization (of Aff.)} \\
 \textit{University of New South Wales}\\Sydney, NSW, Australia\\
 minzhao.lyu@unsw.edu.au}
 \and
 \IEEEauthorblockN{Vijay Sivaraman}
 \IEEEauthorblockA{%\textit{dept. name of organization (of Aff.)} \\
 \textit{University of New South Wales}\\Sydney, NSW, Australia\\
 vijay@unsw.edu.au}
 }

\maketitle

% The default list of authors is too long for headers}
%\renewcommand{\shortauthors}{X.et al.}

\begin{abstract}
Metaverse virtual reality (VR) applications enable users to socialise, work, entertain, and study online with immersive experiences beyond the classic PC-based interactions. While the 360-degree immersion enables users to be fully engaged in a virtual scenario, suboptimal Quality-of-Experience (QoE) like poorly displayed 3D graphics, disruptive loading time, or motion lagging caused by degraded network Quality-of-Service (QoS) can be perceived by users much worse (such as dizziness) than a monitor visualisation. 
This paper empirically measures user QoE of metaverse VR caused by network QoS. Specifically, by focusing on both public social hubs and private user-created events in three popular metaverse VR applications (Rec Room, VRChat and MultiverseVR), we first identify three metrics, including environment freeze level, peripheral content loading time, and control response time, that describe metaverse user experience. By tuning three network QoS parameters (bandwidth, latency, and packet loss), we benchmark each QoE metric's level from excellent to unplayable. Key insights are revealed, such as freeze of metaverse virtual environment is resilient to latency but sensitive to packet loss, and private user-created events demand better network conditions than public social hubs, providing a reference for ISPs to optimise their network QoS for superlative metaverse user experience.
\end{abstract}


\section{Introduction}
The term ``metaverse'', first coined in the science fiction Snow Crash \cite{SnowCrash}, has often been described as a hypothetical iteration of the Internet, supporting virtual environments through personal computing such as virtual reality (VR) headsets. In metaverse applications, users immerse themselves within a continuous and interconnected environment and create shared experiences with others. Significantly, the metaverse differs from traditional networked applications such as gaming, social media and video streaming as it strives to create a highly customisable environment where users can socialise with friends, work \cite{metaEnvironment}, receive education \cite{MetaEducation}, take healthcare service \cite{MetaHealthCare}, and play without being confined by the bounds of distance, ability, and pre-programmed scenarios. 
Over the last ten years, significant investment has flowed into the development of metaverse applications accessed through VR headsets, drawing the attention and resources of major tech players such as Microsoft, Meta, Nvidia and Apple. %The development of the metaverse has also spurred directed investment into building high-performance cloud computing with Nvidia’s Omniverse platform. 
As predicted by McKinsey \& Company, the metaverse market is likely to grow to \$124Bn in 2024 and to exceed \$5Tn by 2030 \cite{MetaMarket}. 

The metaverse provides an immersive experience to its users through virtual and augmented reality (VR) headsets. To this end, a poor user experience in metaverse delivered by VR headsets will have a more drastic impact on user's physical well-beings (\eg causing dizziness and nausea \cite{chattha2020motion}) compared to other multimedia applications such as online gaming and video streaming. Prior works have studied the impact of network quality conditions (QoS) on online gaming. For example, the authors of \cite{CSGO} highlight how latency in a first-person shooting game can affect user experience. The work described in \cite{madanapalli2021reclive} demonstrates how slow download speeds can cause jitters in video streaming applications. However, no prior work has systematically understood the impact of degraded network QoS on user experience for metaverse VR applications.

As highlighted in a recent work \cite{MetaVRadar}, metaverse VR applications encompass a complex set of distinct user activities, each with unique network traffic characteristics. This complexity makes it inherently more intricate to understand their network demands for superlative user experience compared to other applications.
Specifically, as major activities in a metaverse, users can interact with others in a centralised socialisation area called Main Hub (or \textbf{public} social hub in this paper) or visit private events, such as meetups and parties, created and customised by other users. This major type of user activity is named Separate User-created Event (SUE) by M. Lyu \textit{et al.} \cite{MetaVRadar}, or \textbf{private} user-created events in this paper. The two major metaverse activity types possess distinct characteristics in network communications, necessitating our analysis of their user experience as impacted by network QoS.
Other activity types categorised in \cite{MetaVRadar} like trading digital assets or staying at home space impose negligible network demands, thus, are not discussed in this paper.

For Internet Service Providers (ISPs), optimising network service QoS to ensure user experience (QoE) of emerging metaverse VR applications cannot be overstated. However, a significant research gap exists in understanding the relationship between network QoS and metaverse QoE. This gap hinders ISPs from optimising their networks cost-effectively to meet the unique network demands of metaverse VR sessions through typical traffic engineering techniques like priority queues, network slices, and network APIs. 

This paper aims to provide ISPs with a reference for optimising their network service quality for superlative metaverse user experience. We empirically analyse how metaverse user QoE is affected by different (degradation) levels of network QoS. Three popular metaverse applications (Rec Room, VRChat, and MultiverseVR) available on a dominant VR platform, \ie Meta's Oculus, that have been well studied in prior works \cite{MetaVRadar,ReadyForMetaIMC} are analysed in this paper. %Given that similar conclusions have been drawn from all three metaverse applications studied, we will focus on discussing the insights from the most popular application, Rec Room.

Our contributions can be summarised as two-fold.

\begin{itemize}
\item In the \textbf{first contribution} (in \S\ref{sec:back}), with our first-hand experience in metaverse VR applications in our controlled lab environment, we systematically formalise three QoE metrics, including ``Freeze'', ``Content Loading'', and ``Control Responsiveness'' that collectively describe metaverse VR user experience. We use the ACR scale, which is commonly used to describe the user-perceived experience, to assess our proposed QoE metrics into five ratings from ``Excellent'', ``Good'', ``Fair'', ``Poor'', and ``Bad''.
 
\item Our \textbf{second contribution} (in \S\ref{sec:experience}) evaluates the user experience (described by our three QoE metrics) in two major metaverse events, namely public social hubs and private user-created events, that are served by various constrained network conditions including limited bandwidth, packet loss rate, and network latency. Key insights are obtained such as all QoE metrics, particular ``Freeze'', exhibit greater resilience to increased latency than low bandwidth and high packet loss rate; and public social hubs are more insensitive to degraded network QoS than private user-created events.
We also suggest minimum network service quality (QoS) to be guaranteed by ISPs for five levels of overall metaverse user experience from ``'Excellent'' to ``Bad''.

\end{itemize}

\section{Categorising Metaverse User Experience}
\label{sec:back}

In this section, by conducting lab experiments on representative metaverse applications on the Oculus VR platform with controlled network conditions (\S\ref{sec:ifs}), we discuss experience issues that can be faced by users (\S\ref{sec:userExperienceIssue}) which are captured by our formally articulated QoE metrics (\S\ref{sec:categorise}).

\subsection{Measurement Setup}\label{sec:ifs}
\begin{figure}[!t]
 \vspace{2mm}
	\centering
	\includegraphics[width=0.99\linewidth]{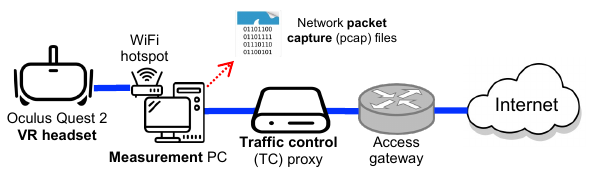}
	\caption{Our measurement setup in controlled lab environment.}
	\label{fig:setup}
\end{figure}

We configure our lab setup (visually shown in Fig.~\ref{fig:setup}) that enables us to measure network traffic characteristics of metaverse VR sessions while enforcing typical quality-of-service (QoS) constraints on network conditions.
During our experiments, we used a standalone Meta's Oculus Quest 2 VR headset connected via WiFi hotspot on our measurement PC to Metaverse service providers on the Internet.
A traffic control proxy (running Linux OS) is connected between the measurement PC and the Internet so that we can quantitatively enforce network constraints, including bandwidth, packet drop rate, and latency. 

Network traffic exchanged between the VR headset and the Internet is captured as PCAP files on the measurement PC. In addition, to correlate the network traffic characteristics with user experience, we keep a user log for the session states, time intervals, and user-perceived experience levels. 
Three popular metaverse applications (\ie Rec Room \cite{Recroom}, VRChat \cite{VRChat} and MultiverseVR \cite{multiverse}) on Oculus VR platforms that have been studied in prior research works \cite{ReadyForMetaIMC, MetaVRadar} are used in our experiments. For brevity, we discuss our insights obtained for Rec Room throughout the rest of this paper as similar conclusions are drawn for all studied metaverse applications.

\begin{figure}[!t]
 \vspace{2mm}
	\centering
	\mbox{
		\hspace{-3mm}
		\subfigure[Long freeze of virtual environment (MultiverseVR).]{
			{\includegraphics[width=0.48\textwidth]{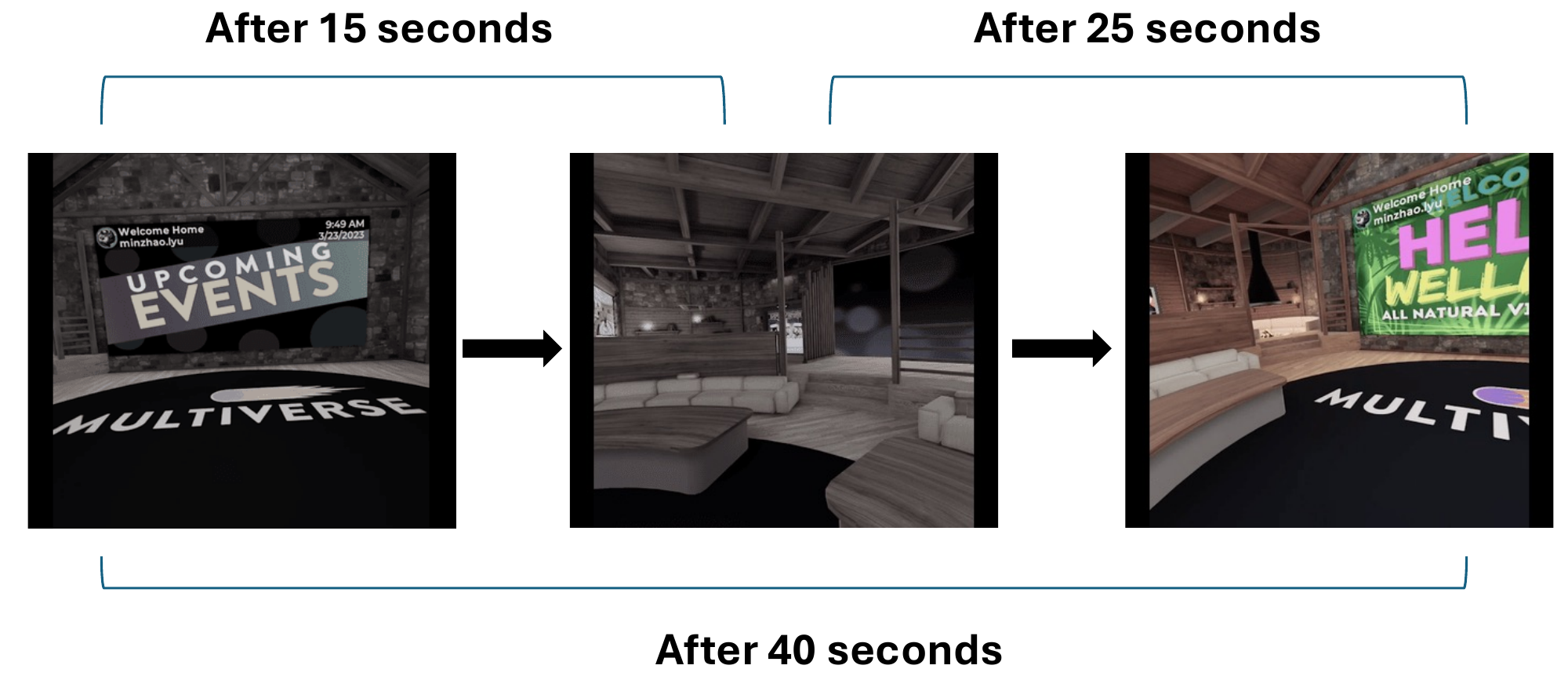}}\quad
			\label{fig:grey}
		}
         }
         \mbox{
		\hspace{-3mm}
		\subfigure[Excessive peripheral content loading times (Rec Room and VRChat).]{
			{\includegraphics[width=0.49\textwidth]{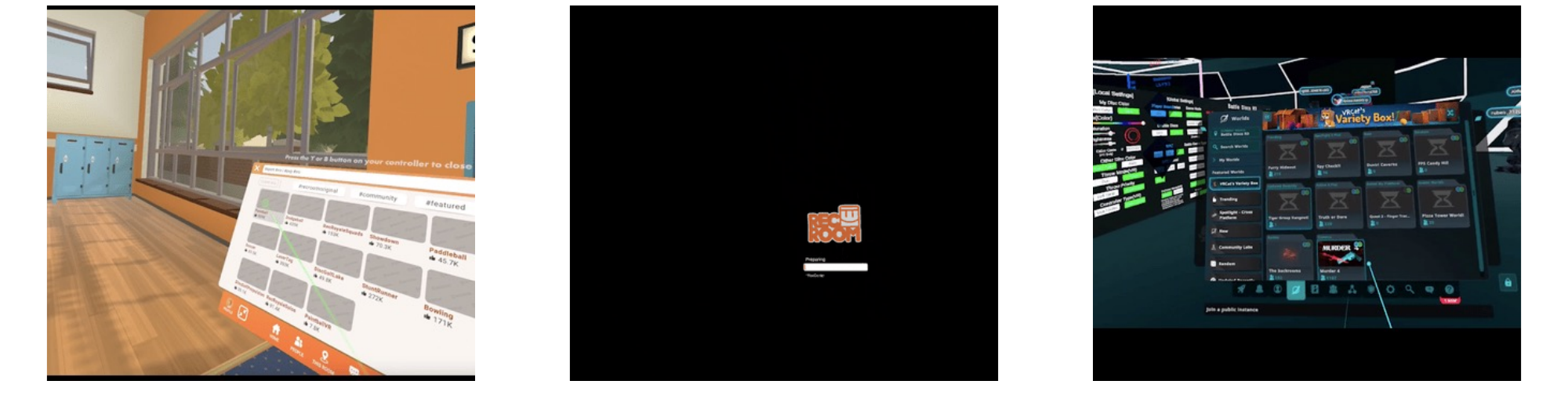}}\quad
			\label{fig:store}
		}
         }
        \mbox{
		\hspace{-3mm}
		\subfigure[Movement jitters and unexpected position teleportation (Rec Room).]{
			{\includegraphics[width=0.49\textwidth]{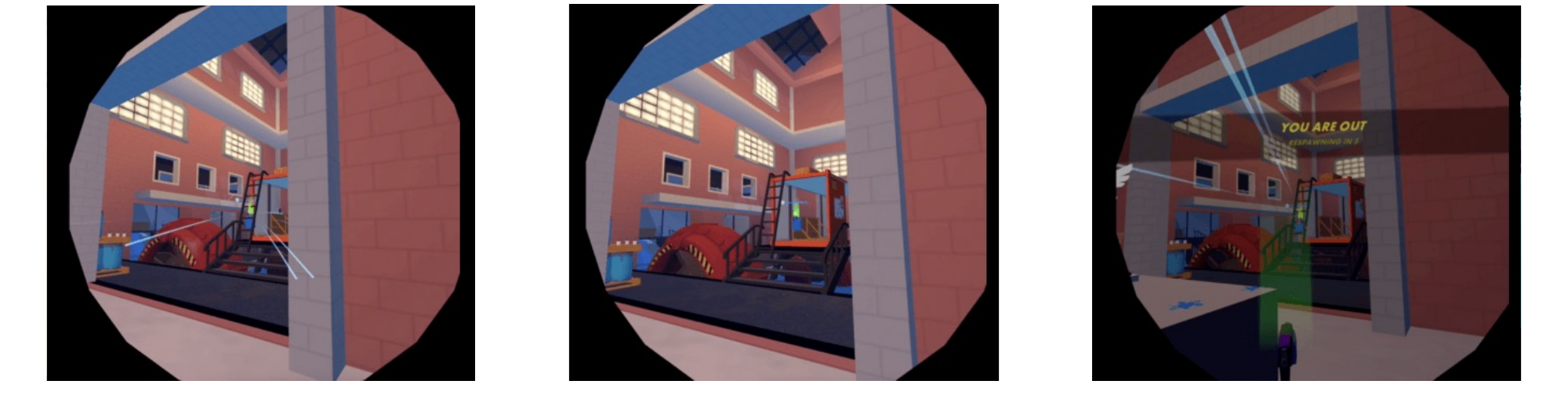}}\quad
			\label{fig:freeze}
		}
	}					
	\caption{Screenshots of representative user experience issues including (a) long freeze of virtual environment, (b) excessive peripheral content loading time, and (c) movement jitters observed by the authors in three popular metaverse VR applications (Rec Room, VRChat and MultiverseVR).}
	\label{fig:experience}
\end{figure}

\subsection{User Experience Issues}\label{sec:userExperienceIssue}

We now discuss two types of experience issue we have experienced in metaverse sessions. Our activities cover all typical scenario types as defined in \textit{MetaVRadar} \cite{MetaVRadar} such as public social hub (MH) and private user-created events (SUE) where users socialise in a public domain or private group, respectively.  

\begin{table*}[t!]
 \vspace{2mm}
 \small
	\caption{Our three metaverse VR user experience metrics and their five rating levels with ACR scale.}
	\label{tab:experience}
	\begin{tabularx}{\linewidth}{|p{1.41cm}|p{1.9cm}|p{2.5cm}|p{3cm}|p{3.2cm}|p{3.5cm}|}
		\hline
		&
		\cellcolor{green}\textbf{Excellent} &
		\cellcolor{green!25}\textbf{Good} &
		\cellcolor{yellow!75}\textbf{Fair}  &
		\cellcolor{orange!75}\textbf{Poor} &
		\cellcolor{red!75}\textbf{Bad} \\ \hline

		\textbf{Freeze} &
		\cellcolor{green}Run smoothly &
		\cellcolor{green!25}Occasional stutters, negligible impact &
		\cellcolor{yellow!75}Frequent jitter, disruptive user experience &
		\cellcolor{orange!75}Almost Unplayable due to severe jittering causing graphics delays &
		\cellcolor{red!75} Unplayable due to extreme, persistent jitter, hindering basic actions \\ \hline

		\textbf{Content Loading} &
		\cellcolor{green}Load instantly &
		\cellcolor{green!25}Load quickly with minimal delay &
		\cellcolor{yellow!75}Moderate loading time with noticeable delay &
		\cellcolor{orange!75}Significant content loading delays evident &
		\cellcolor{red!75}Fail to load \\ \hline

		\textbf{Control Responsiveness} &
		\cellcolor{green}No noticeable delay &
	    \cellcolor{green!25}Slight user action delay, minimal impact &
		\cellcolor{yellow!75}Noticeable action delays cause slight frustration &
		\cellcolor{orange!75}Significant delays between controller and action &
		\cellcolor{red!75}No response between controller and action \\ \hline
		
	\end{tabularx}
\end{table*}

The \textbf{first type} of experience issues we have faced is the freeze of virtual environmental graphics and inconsistency in loading peripheral event content. Fig.~\ref{fig:grey} provides a visual example of the freeze in the virtual environment in one MultiverseVR session when we enter our private home space. As seen from the left to the right of Fig.~\ref{fig:grey}, it took around 40 seconds for the colour of the virtual environment to turn from starkly grey to normal. 
An example of peripheral content loading inconsistency is shown in Fig.~\ref{fig:store}. The two screenshots from the left show partially loaded items/contents in our Rec Room sessions. The leftmost image has a portable user panel fully loaded without the peripheral item, while the scenario captured in the middle screenshot has an environment loaded without items in the menu. In the VRChat session captured by the rightmost screenshot, we saw partially loaded items in the front panel, which stopped us from further interacting within the event.

The \textbf{second type} of experience issues are the fluidity of movements (\ie freezing and teleportation). For example, when we were shooting a basketball in the Rec Room's public social hub (the middle screenshot in Fig.~\ref{fig:store}) with poor network conditions, we saw noticeable jitter with the movements of the ball and players around us as they skip around the court. Another visual example of movement freezing and teleportation is given in Fig.~\ref{fig:freeze} when we play a shooting game in a private event. This issue becomes severe when more players are around us.
Notably, this type of issue can be unacceptable for users wearing VR headsets as the experience is immersive. Users can feel mentally and physically unwell, such as dizziness and nausea when the entire surrounding ``virtual environment'' begins to flicker and tremble, or their own body (movement) control becomes inconsistently reflected in the 3D world.

\subsection{Metaverse VR User Experience (QoE) Metrics}\label{sec:categorise}
We now define three metrics systematically describing the user experience issues discussed above. The three metrics, namely freeze, content loading, and control responsiveness, are measured using the Absolute Category Rating \cite{ACR} (ACR) scale that is commonly used to index user experience of various multimedia applications \cite{ACRREF}.

\subsubsection{Freeze of Metaverse Virtual Environment} The first metric, ``\textbf{Freeze}'', refers to the momentary halt of visuals in the metaverse virtual environment. Specifically, ``freeze'' is characterized by a temporary but noticeable suspension of the virtual environment and can vary in severity from minor stutters to complete halts. This metric significantly impacts experience as the metaverse virtual world encompasses the user's field of view, and even minor disturbances in fluidity can cause nausea or head spinning. A consistent freeze can lead to a loss of presence as the user is taken away from being immersed in the virtual environments delivered by the VR headsets. Therefore, using the ACR scale, we define an excellent experience in freeze when there is no pause in visual interactions and audio communications in a metaverse session, while a bad experience in freeze with extreme jitter, unbearable screen shaking, etc., can lead to nausea and headache. The middle levels, including good, fair and poor, are specified in the first row of Table~\ref{tab:experience}.

\subsubsection{Loading of Peripheral Content} We define our second metric as ``\textbf{Content Loading}'' to describe how fast the peripheral content is loaded when metaverse users interact with the virtual environment. Unlike online games that pre-download all peripheral contents with the application software, as discussed in \cite{MetaVRadar, gamingQoE}, the virtual contents such as environment layout, assets, decorations and textures in metaverse sessions are often designed or placed by other users and fetched from content servers to the client VR headset in real-time.
In poor network conditions, incomplete or slow loading of contents can lead to a disjointed user experience and stop users from socialising with others due to missing items. Using the ACR scale, we define five levels of user experience in content loading, from excellent, good, fair, to poor and bad. The detailed qualitative definition of each experience level is provided in the second row of Table~\ref{tab:experience}.

\subsubsection{Synchronization of User Action/Motion} Our third metaverse user experience (QoE) metric is ``\textbf{Control Responsiveness}'', which indicates the duration taken to reflect the users' body motion and action in the virtual environment. A fast and consistent control response is crucial for a smooth and immersive experience, particularly through a VR headset that tracks users' physical body movement and position as inputs. A highly delayed control response (visually) perceived by users in an immersive virtual environment can cause frustration and physical discomfort. As shown in the last row in Table~\ref{tab:experience}, using the ACR scale, we define five levels of user experience for control responsiveness, from excellent (no noticeable delay) to a bad experience when extremely unplayable (high) response delay is perceived by users through VR headsets.

\section{Network QoS Conditions Impacting Metaverse VR User Experience (QoE)}
\label{sec:experience}
With an intuitive understanding of experience issues that can be perceived by metaverse VR users and formal definition of QoE metrics, in this section, we empirically benchmark the impact of typical network conditions, \ie available bandwidth (\S\ref{sec:bandwidth}), packet loss rate (\S\ref{sec:packetLoss}), latency (\S\ref{sec:latency}) on the QoE metrics (including game loading, freeze, and control responsiveness). Last, we benchmark the minimum requirements of the three network QoS metrics to support overall metaverse user experience at five levels from excellent to bad (\S\ref{sec:multipleFactor}).

\subsection{Available Bandwidth}\label{sec:bandwidth}
We first discuss how bandwidth limitation impacts the user experience in Rec Room sessions. As discussed in \S\ref{sec:ifs}, we use Linux traffic control (TC) functions on the traffic control proxy in Fig.~\ref{fig:setup} to adjust the available bandwidth to our VR headset. Packet streams exchanged between the VR headset and metaverse servers on the Internet are captured by our measurement PC for post-hoc analysis. Our controlled experiments are conducted for both public social hubs and private user-created events.

\subsubsection{Public Social Hub} 
\begin{figure}[!t]
\vspace{2mm}
	\includegraphics[width=0.49\textwidth]{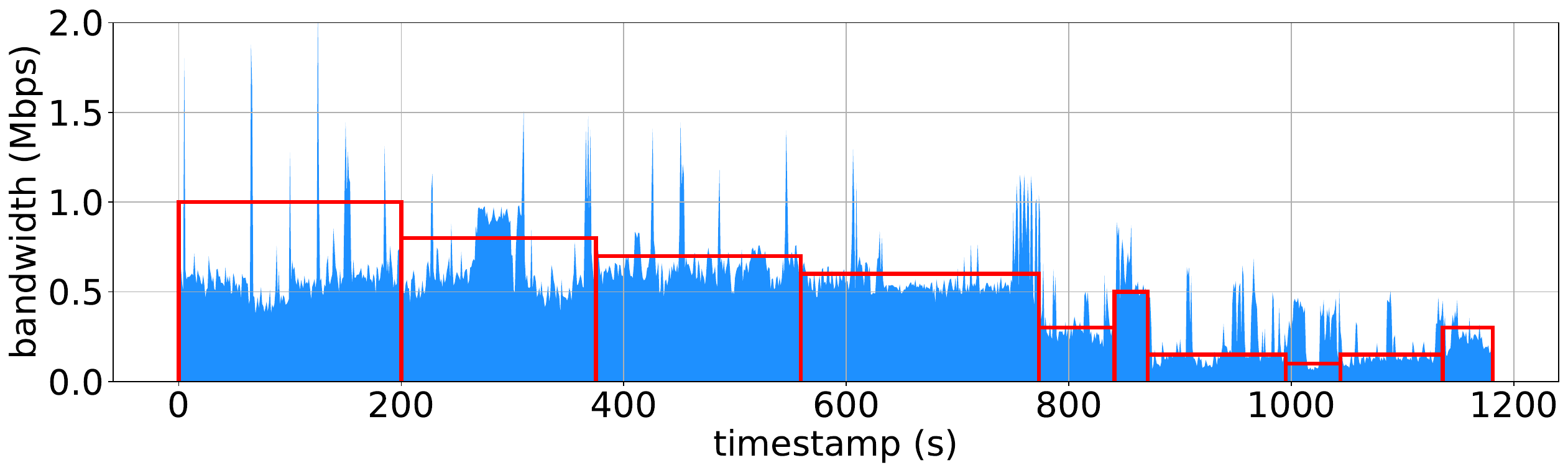}
	\caption{Bandwidth usage by one public social hub metaverse session with bandwidth constraints shown as red frames.}
	\label{fig:bandwidthmh}
\end{figure}

We gradually reduce the available bandwidth from unlimited to 1Mbps, 700kbps, and 600kbps till 150kbps when the session starts to crash. One representative example of the actual bandwidth consumed by the metaverse session through our VR headset with constraints is shown as the time-series plot in Fig.~\ref{fig:bandwidthmh}. 

\begin{table}[!t]
	\caption{User experience (color-coded as \textbf{\color{green}Excellent}, \textbf{\color{green!25}Good}, \textbf{\color{yellow!75}Fair}, \textbf{\color{orange!75}Poor} and \textbf{\color{red!75}Bad}) in Public Social Hub with Bandwidth Constraints.}
	\label{tab:bandwithmh}
	\resizebox{0.99\columnwidth}{!}{\begin{tabular}{|l|l|l|l|l|l|l|l|}
			\hline
			&  \textbf{700kbps}  &  \textbf{600kbps} & \textbf{500kbps}   & \textbf{400kbps}   & \textbf{300kbps}  &  \textbf{200kbps}   \\ \hline
			\textbf{Freeze}                & \cellcolor{green} Excel. &  \cellcolor{green!25} Good & \cellcolor{yellow!75} Fair & \cellcolor{orange!75} Poor & \cellcolor{orange!75} Poor & \cellcolor{red!75} Bad   \\ \hline

   \textbf{Content loading}          & \cellcolor{green} Excel.  & \cellcolor{green} Excel. & \cellcolor{green!25} Good & \cellcolor{yellow!75} Fair & \cellcolor{red!75} Bad & \cellcolor{red!75} Bad   \\ \hline

     \textbf{Control Resp.} & \cellcolor{green} Excel. & \cellcolor{green!25} Good & \cellcolor{green!25} Good & \cellcolor{green!25} Good & \cellcolor{red!75} Bad & \cellcolor{red!75} Bad \\ \hline

	\end{tabular}}
\end{table}

After repeating multiple (\eg three) times for the same set of bandwidth limitations, we have observed that with unlimited, 1Mbps and 700kbps limits, the user experience in public social hub is visually smooth, exhibiting immediate control response and negligible loading delay for peripheral contents. Therefore, we mark the three QoE metrics (freeze, loading time, and control responsiveness) under 700kbps bandwidth limitation as an excellent level, as shown in the first column of Table~\ref{tab:bandwithmh}. 
With the 600kbps bandwidth limitation being enforced, we have noticed slight graphical jitter and response delay in both freeze and control responsiveness while the content loading is still excellent. 
When we further reduce the available bandwidth to lower levels, the user experience in control responsiveness exhibits quite robust behaviour compared to the other two metrics. We have a ``good'' rating for these metrics even with only 300kbps available bandwidth. Notably, ``freeze'' and ``content loading'' drop to ``Fair'' and ``Poor'' ratings with 300kbps.

\begin{figure}[!t]
   \vspace{2mm}
	\includegraphics[width=0.49\textwidth]{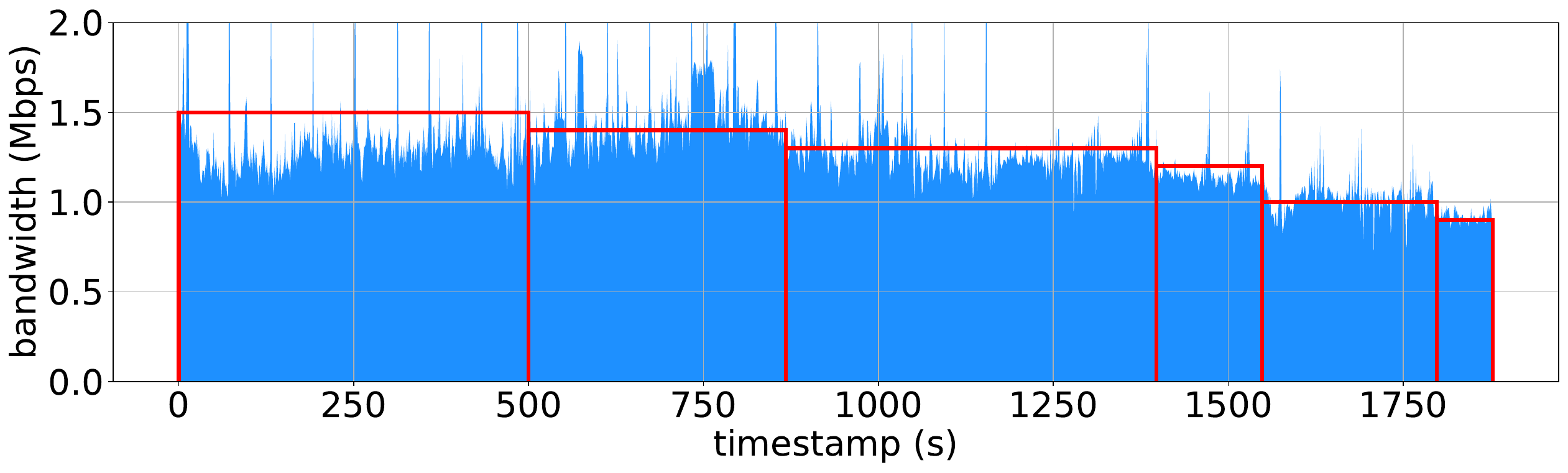}
	\caption{Bandwidth usage by one private user-created metaverse event with bandwidth constraints shown as red frames.}
	\label{fig:bandwidthsue}
\end{figure}

\begin{table}[!t]
	\caption{User experience (\textbf{\color{green}Excellent}, \textbf{\color{green!25}Good}, \textbf{\color{yellow!75}Fair}, \textbf{\color{orange!75}Poor} and \textbf{\color{red!75}Bad}) in Private User-created events with Bandwidth Constraints.}
	\label{tab:bandwidthsue}
	\resizebox{0.99\columnwidth}{!}{\begin{tabular}{|l|l|l|l|l|l|l|l|}
			\hline
			&  \textbf{1.5Mbps}  &  \textbf{1.3Mbps}  & \textbf{1.2Mbps}    &   \textbf{1.0Mbps}    &  \textbf{900kbps}   \\ \hline

		\textbf{Freeze}                &  \cellcolor{green} Excellent &  \cellcolor{green!25} Good & \cellcolor{yellow!75} Fair & \cellcolor{orange!75} Poor & \cellcolor{red!75} Bad \\ \hline
		\textbf{Content Loading}          & \cellcolor{green} Excellent  & \cellcolor{green} Excellent & \cellcolor{green!25} Good & \cellcolor{yellow!75} Fair & \cellcolor{red!75} Bad \\ \hline
		
\textbf{Control Resp.} & \cellcolor{green} Excellent & \cellcolor{green} Excellent & \cellcolor{green!25} Good & \cellcolor{green!25} Good & \cellcolor{red!75} Bad  \\ \hline
	\end{tabular}}
\end{table}

\subsubsection{Private User-created Events}
Private events (\eg group party, ceremony, classroom and meetups) that are created, designed and decorated by metaverse users have much higher bandwidth demands for decent user experience than public social hubs. We tune the bandwidth limitation from unlimited, 1.5Mbps, 1.3Mbps, 1.2Mbps, 1.0Mbps, to 900kbps when the events become unplayable. An example of bandwidth usage with enforced bandwidth limitations is shown as a time-series plot in Fig.~\ref{fig:bandwidthsue}.

From Table~\ref{tab:bandwidthsue}, where we show the QoE levels for the three metrics as observed from our repeated experiments, it is obvious that 1.5Mbps and 1.3Mbps are required to achieve ``Excellent'' and ``Good'' experience for all three metrics, respectively. With 1.0Mbps available bandwidth, ``freeze'' becomes ``Poor'' and ``content loading'' becomes ``Fair'', indicating a quite undesirable overall user experience. 
Importantly, with 900kbps available bandwidth to the VR headset, all three user experience metrics are rated as "Bad" in private user-created events. In comparison, a public social hub only requires 700kbps for an optimal metaverse user experience level.

\subsubsection{\textbf{Key Takeaways}}
In summary, control responsiveness has the least demand on available bandwidth for public social hubs in the metaverse. Content loading is also not very sensitive to limited bandwidth as most peripherals in public social hubs are pre-installed on local VR headsets instead of fetching from remote servers in real-time. The smoothness of virtual environment has the most bandwidth demands (\eg 600kbps for ``Good'' rating).

In private user-created events, among the three QoE metrics, freeze is still more sensitive to limited bandwidth than content loading and control responsiveness, consistent with our observations in public social hubs. However, we have identified significantly higher bandwidth demands in private user-created events than in public social hubs, which differs from the other two network constraints (packet loss rate and latency) that will be discussed next.

\begin{figure}[!t]
 \vspace{2mm}
	\includegraphics[width=0.49\textwidth]{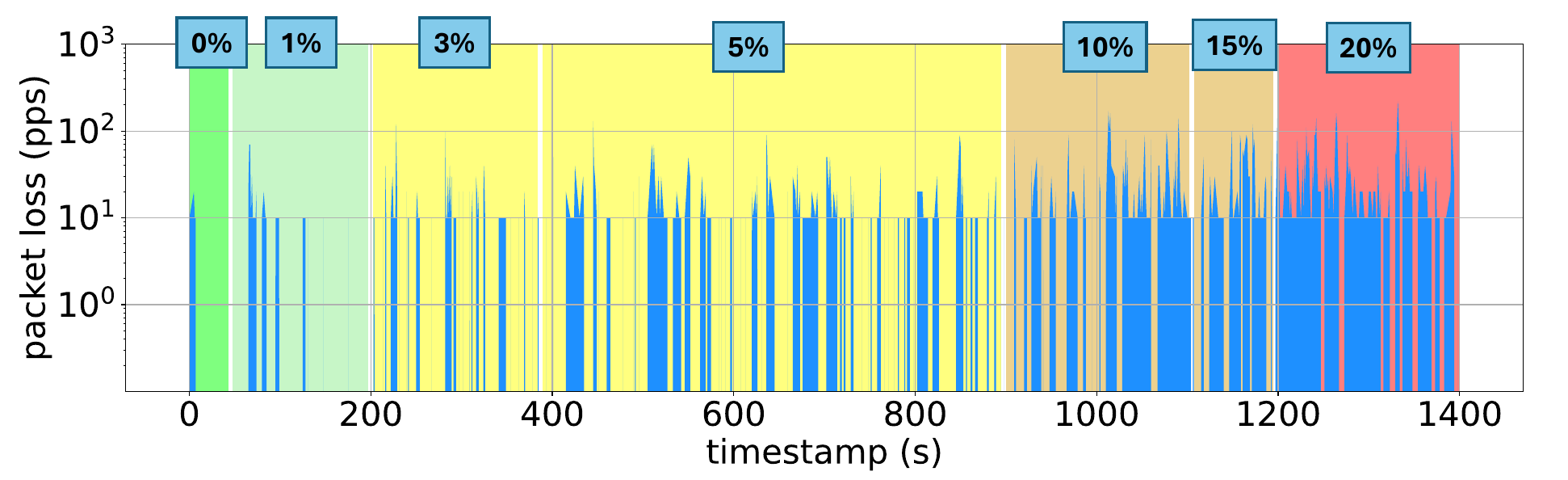}
	\caption{Number of packet loss per second in one public social hub metaverse session with various packet loss rates.}
	\label{fig:lossmh}
\end{figure}

\begin{table}[!t]
        \caption{User experience (color-coded as \textbf{\color{green}Excellent}, \textbf{\color{green!25}Good}, \textbf{\color{yellow!75}Fair}, \textbf{\color{orange!75}Poor} and \textbf{\color{red!75}Bad}) in public social hub with packet loss rate.}
	\label{tab:lossmh}
	\resizebox{0.99\columnwidth}{!}{\begin{tabular}{|l|l|l|l|l|l|l|l|}
			\hline
			& \textbf{0\%}  &  \textbf{1\%} &  \textbf{3\%}   &  \textbf{5\%}   &  \textbf{10\%}  & \textbf{15\%}  & \textbf{20\%}  \\ \hline
			
   \textbf{Freeze}                & \cellcolor{green} Excellent & \cellcolor{green!25} Good & \cellcolor{yellow!75} Fair  &\cellcolor{yellow!75} Fair  & \cellcolor{orange!75} Poor & \cellcolor{red!75} Bad  & \cellcolor{red!75} Bad \\ \hline
			
			\textbf{Content Loading}          & \cellcolor{green} Excellent  & \cellcolor{green} Excel. & \cellcolor{green!25} Good & \cellcolor{green!25} Good& \cellcolor{yellow!75} Fair  &\cellcolor{orange!75} Poor   & \cellcolor{red!75} Bad \\ \hline

			\textbf{Control Resp.} & \cellcolor{green} Excellent &\cellcolor{green} Excel. &  \cellcolor{green!25} Good&   \cellcolor{green!25} Good & \cellcolor{yellow!75} Fair & \cellcolor{red!75} Bad & \cellcolor{red!75} Bad \\ \hline
	\end{tabular}}
\end{table}

\subsection{Packet Loss Rate}\label{sec:packetLoss}
We now look into the user experience issues (described by our three QoE metrics) caused by the packet loss rate. Similar to our experimental process in \S\ref{sec:bandwidth}, we now tune the packet loss rate (in percentage) gradually from 0\% (no loss) to 20\% when all metaverse sessions become unplayable. In addition to the perceived experience of our three defined metrics, we also compute the number of lost packets per second in real-time, which is proportional (as defined by the loss rate) to the packet rate of the current metaverse session.

\subsubsection{Public Social Hub}
The user experience and the number of lost packets per second in the public social hub are given in Table~\ref{tab:lossmh} and Fig.~\ref{fig:lossmh}, respectively. As we can see from Fig.~\ref{fig:lossmh}, when we introduce a higher packet loss rate (in percentage), the number of lost packets becomes larger.

Under ideal conditions with no packet loss, we could not perceive any noticeable content loading/control delay and environment freeze, leading to an excellent rating for all three experience metrics, shown in the first column of Table~\ref{tab:lossmh}. When the packet loss rate increases to 1\%, we start experiencing slight inconsistency when moving around the virtual environment while the other two metrics remain unaffected, resulting in the rating for ``Freeze'' dropping from ``Excellent'' to ``Good''. 
With 3\% and 5\% drop rates, we perceive noticeable degradation in the environment freeze and a slight delay in peripheral content loading and control response, resulting in ``Good'' ratings for ``Content Loading'' and `` Control Responsiveness'' and ``Fair'' for ``Freeze''.

A drastic increase in environment freezes and screen shakes are observed when the packet loss rate becomes 10\%. One of the authors of this experiment started feeling dizzy due to frequent screen shakes when moving in the virtual environment. Also, noticeable delays in loading peripheral content and control response are observed with this setting.
Higher packet loss rates than 10\% make the user experience unplayable (\ie ``Bad'' in one or more metrics), as shown in the last two columns of Table~\ref{tab:lossmh}. 

\begin{figure}[!t]
 \vspace{2mm}
	\includegraphics[width=0.49\textwidth]{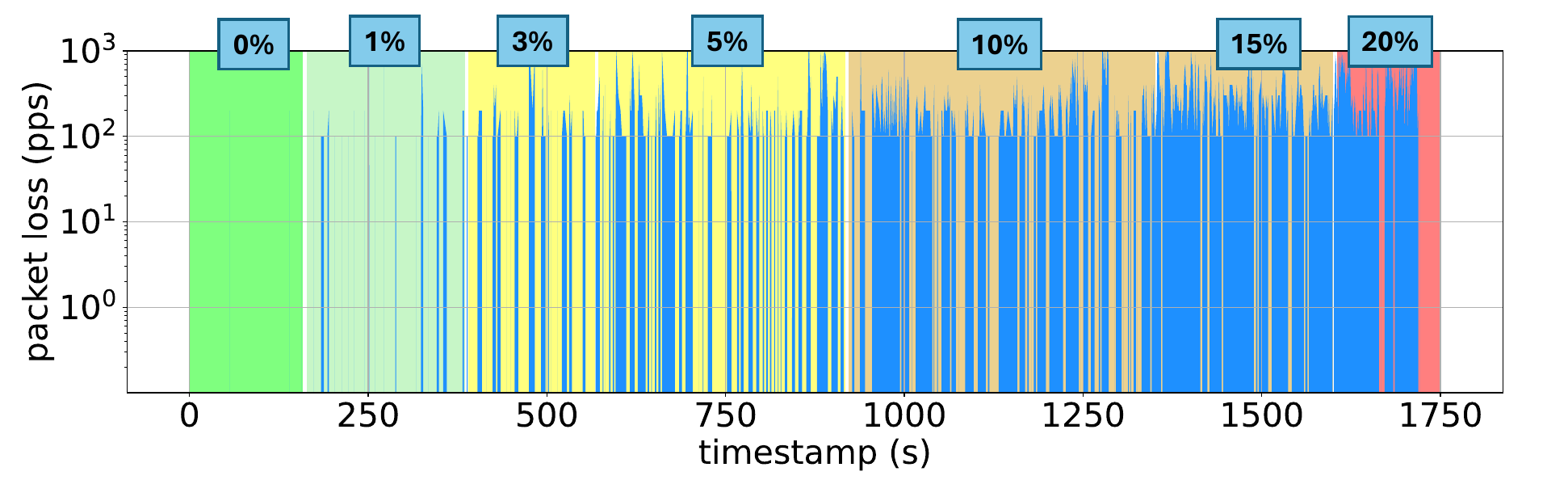}
	\caption{Number of packet loss per second in one private user-created event with various packet loss rates.}
	\label{fig:losssue}
\end{figure}

\begin{table}[!t]
	\caption{User experience (color-coded as \textbf{\color{green}Excellent}, \textbf{\color{green!25}Good}, \textbf{\color{yellow!75}Fair}, \textbf{\color{orange!75}Poor} and \textbf{\color{red!75}Bad}) in private user-created events with packet loss rate.}
	\label{tab:losssue}
	\resizebox{0.99\columnwidth}{!}{\begin{tabular}{|l|l|l|l|l|l|l|l|}
			\hline
			& \textbf{0\%}  &  \textbf{1\%} &  \textbf{3\%}   & \textbf{5\%}   & \textbf{10\%}  & \textbf{15\%}  & \textbf{20\%}  \\ \hline
			\textbf{Freeze}                & \cellcolor{green} Excellent & \cellcolor{green!25} Good & \cellcolor{yellow!75} Fair  &\cellcolor{yellow!75} Fair  & \cellcolor{orange!75} Poor & \cellcolor{red!75} Bad  & \cellcolor{red!75} Bad \\ \hline
			
			\textbf{Content Loading}          & \cellcolor{green} Excellent  & \cellcolor{green} Excel. & \cellcolor{green!25} Good & \cellcolor{green!25} Good& \cellcolor{yellow!75} Fair  &\cellcolor{orange!75} Poor   & \cellcolor{red!75} Bad \\ \hline

			\textbf{Control Resp.} & \cellcolor{green} Excellent &\cellcolor{green} Excel. &  \cellcolor{green!25} Good&   \cellcolor{yellow!75} Fair & \cellcolor{yellow!75} Fair & \cellcolor{red!75} Bad & \cellcolor{red!75} Bad \\ \hline
	\end{tabular}}
\end{table}

\subsubsection{Private User-created Events}
Unlike the conclusions we draw for available bandwidth, private user-created events exhibit similar sensitivity to packet drops compared to metaverse public social hubs.
Fig.~\ref{fig:losssue} shows the gradually increased number of dropped packets that is consistent with our configured packet loss rate. Table~\ref{tab:losssue} illustrates the ratings of our three user experience metrics. 
When the packet drop rate increases to 1\%, content loading and control responsiveness are not impacted, while the freeze metric drops from ``Excellent'' to ``Good''. Compared to our observations for public social hubs in Table~\ref{tab:lossmh}, according to Table~\ref{tab:losssue}, exact the same ratings exhibit for all three metrics when the packet drop rate gradually jumps to 20\%. One exception is that the control responsiveness becomes ``Fair'' with a 5\% packet drop rate in private user-created events while remaining ``Good'' in public social hubs. We guess the reason behind is that, public social hubs that are operated by the metaverse operators have user motion predictions implemented to compensate a certain level of packet loss, while such prediction can be hardly implemented by individual users who create their own events.

\subsubsection{\textbf{Key Takeaways}}
We highlight three key observations for packet loss rate.
First, public social hubs and private user-created events exhibit similar sensitivity to packet loss.
Second, freeze, which indicates the smoothness of environmental changes, is the most sensitive metric to packet loss in both metaverse events.
Third, public social hubs have slightly better tolerance for packet loss in control responsiveness than private user-created events.

\subsection{Latency}\label{sec:latency}
We now report our empirical analysis on how network latency impacts metaverse user experience in public social hubs and private user-created events.

\subsubsection{Public Social Hub}
\begin{figure}[t!]
 \vspace{2mm}
	\includegraphics[width=0.49\textwidth]{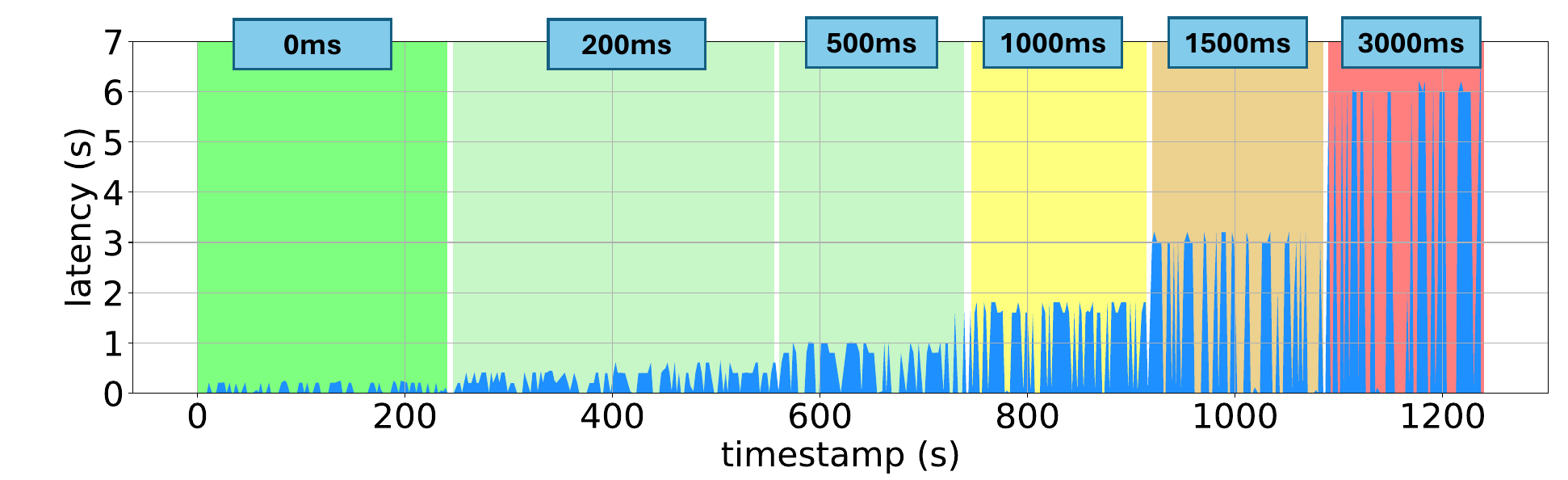}
	\caption{Real-time latency between VR headset and (Rec Room) metaverse server in one public social hub with various enforced latency levels.}
	\label{fig:latencymh}
\end{figure}

We enforce various levels of network latency (from 0 milliseconds to 5 seconds in seven steps as shown in Table~\ref{tab:latencymh}) between the VR headset and metaverse server(s) when experimenting in public social hubs. The measured real-time latency is visually shown in Fig.~\ref{fig:latencymh}, proving the precision of our setup.

The user experience before reaching 300ms remains at an excellent level for all three metrics, which is quite surprising as a latency above 100ms is often considered detrimental to user experience in many interactive multimedia applications such as gaming \cite{madanapalli2022know} and cloud gaming \cite{ReadyForMetaIMC}. 
The ``Content Loading'' drops to ``Good'' with 300ms latency due to a slight delay in peripheral content loading while the other two metrics remain ``Excellent''.

When we further increase the latency, the ``Freeze'' metric stays at ``Excellent'' level until the latency becomes 3 seconds, whereas the other two metrics become ``Fair'' (\ie noticeable degradation) with 500ms latency. We believe that the superior performance in synchronizing with the virtual environment of public social hubs is due to two reasons: first, the environment content is pre-installed on the VR headset, and second, metaverse VR applications adopt motion prediction algorithms to compensate for high latency scenarios.

\begin{table}[t!]
	\caption{User experience (color-coded as \textbf{\color{green}Excellent}, \textbf{\color{green!25}Good}, \textbf{\color{yellow!75}Fair}, \textbf{\color{orange!75}Poor} and \textbf{\color{red!75}Bad}) in public social hub with network latency.}
	\label{tab:latencymh}
	\resizebox{0.99\columnwidth}{!}{\begin{tabular}{|l|l|l|l|l|l|l|l|}
			\hline
			& \textbf{0ms}  &  \textbf{100ms} &  \textbf{300ms}   & \textbf{500ms}   &  \textbf{1s}  & \textbf{3s} \\ \hline
			\textbf{Freeze}                & \cellcolor{green} Excellent &  \cellcolor{green} Excel. & \cellcolor{green} Excel. &\cellcolor{green} Excel. & \cellcolor{green} Excel. & \cellcolor{red!75} Bad \\ \hline
			
			\textbf{Content Loading}          & \cellcolor{green} Excellent  & \cellcolor{green} Excel. & \cellcolor{green!25} Good & \cellcolor{yellow!75} Fair & \cellcolor{red!75} Bad & \cellcolor{red!75} Bad  \\ \hline

			\textbf{Control Resp.} & \cellcolor{green} Excellent &\cellcolor{green} Excel. & \cellcolor{green} Excel.&  \cellcolor{yellow!75} Fair & \cellcolor{orange!75} Poor & \cellcolor{red!75} Bad \\ \hline
	\end{tabular}}
\end{table}

\subsubsection{Private User-created Event}
\begin{figure}[t!]
 \vspace{2mm}
		\includegraphics[width=0.49\textwidth]{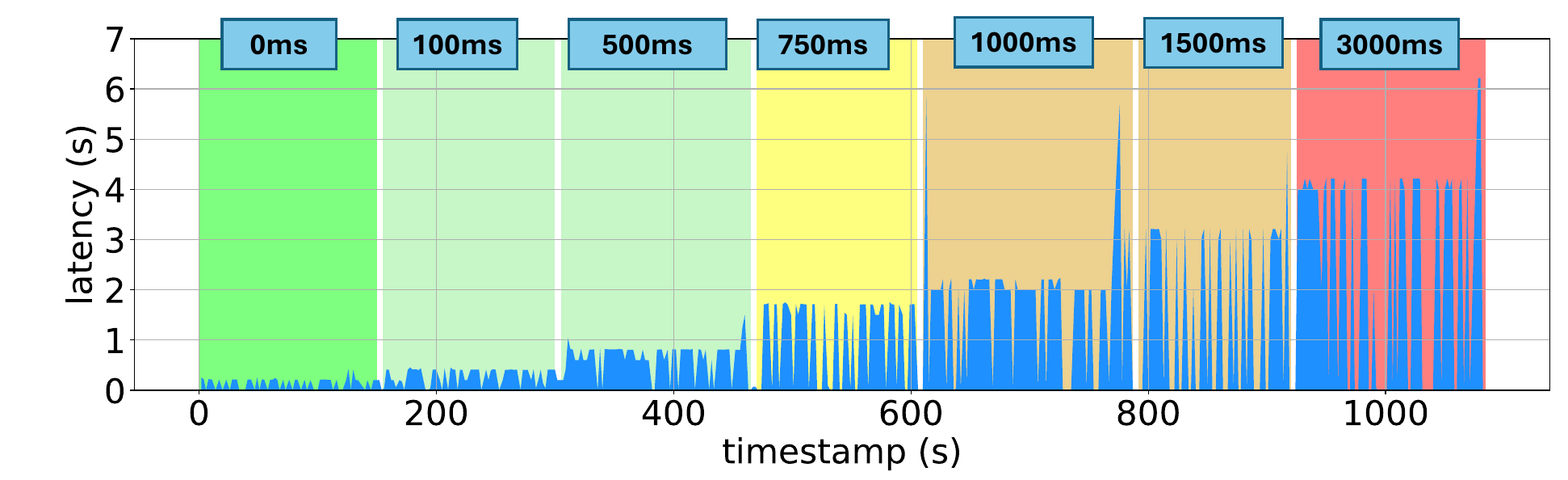}
	\caption{Real-time latency between VR headset and (Rec Room) metaverse server in one private user-created event with various enforced latency levels.}
	\label{fig:latencysue}
\end{figure}

As expected, private user-created events are more sensitive to network latency as most (if not all) event contents are not pre-installed on the VR headset. Our experimental results are summarised in Table~\ref{tab:latencysue}. The time-series plot showing actual latency between the VR headset and Rec Room server(s) is presented in Fig.~\ref{fig:latencysue}.

With a latency of 100 milliseconds for private user-created events, the ``Freeze'' and ``Content Loading'' metrics drop from ``Excellent'' to ``Good'' while the ``Control Responsiveness'' remains ``Excellent''.
When we further increase the latency values from 300ms to 500ms and 1s, the environmental freeze level stays at ``Good'', while the other two metrics drop one level per step.
For high latency scenarios, like our observations for the public social hubs, the metaverse user experience becomes quite terrible regarding peripheral content loading with 1-second latency and completely unplayable with 3-second latency.

\subsubsection{\textbf{Key Takeaways}}
We now discuss three key takeaways from our experiments on network latency.
First, public social hubs are more tolerant to network latency than private user-created events for all three user experience metrics.
Second, peripheral content loading is the most delay-sensitive metric, particularly in private user-created events.
Third, the ``Freeze'' metric that indicates the smoothness of the VR virtual environment exhibits good insensitivity to latency in both public social hubs and private user-created events. This is probably due to the high resilience introduced by motion prediction mechanisms commonly implemented in interactive VR applications.

\subsection{Benchmarking Network QoS Conditions for Five Levels of Overall User QoE Ratings}\label{sec:multipleFactor}
On a real network constrained with poor network QoS, it is unlikely that only bandwidth, packet loss, or latency will be affected individually without the degradation of another network parameter. Therefore, after understanding the user experience impacts caused by a single degraded network QoS factor, we now discuss representative scenarios with multiple degraded network QoS factors.
By varying bandwidth, packet loss rate and latency for public social hubs and private user-created events, we obtained five combinations of network QoS factors for each type of metaverse event that result in ``Excellent'', ``Good'', ``Fair'', ``Poor'' and ``Bad'' ratings in all three user experience metrics, respectively.

\begin{table}[t!]
	\caption{User experience (color-coded as \textbf{\color{green}Excellent}, \textbf{\color{green!25}Good}, \textbf{\color{yellow!75}Fair}, \textbf{\color{orange!75}Poor} and \textbf{\color{red!75}Bad}) in private user-created events with network latency.}
	\label{tab:latencysue}
	\resizebox{0.99\columnwidth}{!}{\begin{tabular}{|l|l|l|l|l|l|l|l|}
			\hline
			&\textbf{0ms}  &   \textbf{100ms} &  \textbf{300ms}   & \textbf{500ms}   & \textbf{1s}  & \textbf{3s} \\ \hline
			\textbf{Freeze}                &\cellcolor{green} Excellent & \cellcolor{green!25} Good  & \cellcolor{green!25} Good  &\cellcolor{green!25} Good  & \cellcolor{green!25} Good & \cellcolor{red!75} Bad  \\ \hline
			
			\textbf{Content Loading}          & \cellcolor{green} Excellent  & \cellcolor{green!25} Good  &\cellcolor{yellow!75} Fair  & \cellcolor{orange!75} Poor &  \cellcolor{red!75} Bad & \cellcolor{red!75} Bad   \\ \hline

			\textbf{Control Resp.} & \cellcolor{green} Excellent &\cellcolor{green} Excel. &  \cellcolor{green!25} Good&  \cellcolor{yellow!75} Fair & \cellcolor{orange!75} Poor & \cellcolor{red!75} Bad  \\ \hline
	\end{tabular}}
\end{table}

Table~\ref{tab:comb1} shows the five combinations of QoS factors that lead to five different user experience ratings in public social hubs. Optimally, a network condition with latency at less or equal to 200ms, an available bandwidth larger or equal to 700kbps, and a 0\% packet loss rate is required for an ``Excellent'' overall user experience in public social hubs. Users may start feeling unwell (\ie a ``Poor'' or even ``Bad'' user experience) when wearing VR headsets with over 1000ms latency, less than 300kbps bandwidth and more than 10\% packet drop rates.

As shown in Table~\ref{tab:comb2}, compared to the five scenarios in public social hubs, we see stricter network QoS requirements for the same user experience level in private user-created events.
For an excellent user experience, almost no extra latency, over 1.5Mbps available bandwidth, and near zero packet loss rate are expected. To avoid poor user experience, a much low latency (\ie 500ms instead of 1000ms) and a much higher bandwidth (\ie 1.1Mbps instead of 300kbps) are required in private user-created events.

\begin{table}[t!]
        \centering
	\caption{Threshold Network QoS Constraints (Bandwidth, Packet Loss and Latency) of Excellent to Bad User Experiences \\in public social hubs.}
 \small
	\label{tab:comb1}
	\resizebox{0.99\columnwidth}{!}{\begin{tabular}{|l|l|l|l|l|l|l|l|}
			\hline
			  & \cellcolor{green} \textbf{Excellent} & \cellcolor{green!25} \textbf{Good}  & \cellcolor{yellow!75} \textbf{Fair} & \cellcolor{orange!75} \textbf{Poor} & \cellcolor{red!75} \textbf{Bad}   \\ \hline
			\textbf{Bandwidth} & 700kbps & 600kbps &  400kbps & 300kbps& 200kbps \\ \hline     
			\textbf{Packet Loss} &  0\% & 3\% &  5\%& 10\%&   20\%\\ \hline
			\textbf{Latency} & 200ms  &  300ms&  500ms &1000ms & 3000ms \\ \hline
	\end{tabular}}
\end{table}

\section{Related Work}
In this section, we first discuss related works on evaluating metaverse user experience from various aspects such as VR hardware configurations, edge computing, and human affective behaviours. We then discuss recent network analysis studies on metaverse VR applications and research works that evaluate the impact of network QoS on user QoE for other applications like video streaming, online gaming, and cloud gaming.

\textbf{Metaverse User Experience:} Metaverse applications on virtual reality (VR) platforms are inherently complex and encompass a broad range of factors that impact experience. Prior studies have focused on user experience impacted by VR headsets' hardware, networking performance, and user interactions \cite{jiang2023joining, liu2023complex, famili2023wi, song2023enable, ahsani2023novel, tsunekiyo2023device, abdelzaher2023challenges, von2023identification}.  Regarding hardware performance, the study in \cite{metahardware} discussed that headset battery depletion occurs 15\% faster under Wi-Fi loading than local execution. Furthermore, it investigated CPU, GPU, and memory usage across a metaverse session, ultimately demonstrating the necessity for all metaverse systems to deliver optimal performance for a consistently positive user experience. Furthermore, the work in \cite{metaedge} analysed the performance of cloud-edge metaverse applications and demonstrates that prefetching content from edge server caches can dramatically increase QoE. The work discussed in \cite{Attention, du2022exploring} optimises resource allocation to the user's current VR focal point, and the work in \cite{comfort} analysed the comfort of virtual reality headset designs for longer user engagement periods and better perceived QoE. For metaverse experience related to user social interactions, the authors of \cite{social} discussed QoE assessment methods used to analyse physiological, human cognitive, and affective behaviour within metaverse applications used to foster positive interactions such as building friendships and working together in games while concurrently mitigating negative experiences, e.g. bullying and discrimination.

\textbf{Metaverse Network Traffic Analysis:} A notable gap persists in measuring user experience from a networking perspective and the network quality factors that underlie metaverse user experience delivered through VR headsets, pioneered by our work in this paper. Some prior work has studied the network traffic characteristics of metaverse applications. Most recently, \textit{MetaVRadar} \cite{55} analysed the network anatomy of metaverse applications and characterised major user activity states. Furthermore, the works \cite{ReadyForMetaIMC, metaflows} measured the network throughput of both the initialization and social interaction stages, end-to-end latency and control response time of a user's actions. They further demonstrate metaverse application latency scales with the number of users leading to potential scalability issues.
Motivated by prior works, we demonstrate the extent to which network conditions impact major metaverse user activities, \ie public social hub and private user-created events.

\begin{table}[t!]
        \centering
	\caption{Threshold Network QoS Constraints (Bandwidth, Packet Loss and Latency) of Excellent to Bad User Experiences \\in private user-created events.}
 \small
	\label{tab:comb2}
	\resizebox{0.99\columnwidth}{!}{\begin{tabular}{|l|l|l|l|l|l|l|l|}
			\hline
			  & \cellcolor{green} \textbf{Excellent} & \cellcolor{green!25} \textbf{Good}  & \cellcolor{yellow!75} \textbf{Fair} & \cellcolor{orange!75} \textbf{Poor} & \cellcolor{red!75} \textbf{Bad}   \\ \hline
			\textbf{Bandwidth} &  1.5Mbps & 1.3Mbps &1.2Mbps & 1.1Mbps&  900kbps \\ \hline     
			\textbf{Packet Loss} &0\% & 3\% & 5\%&  10\%&  20\%\\ \hline
			\textbf{Latency} &0ms  &100ms&  300ms & 500ms &  1500ms \\ \hline
	\end{tabular}}
\end{table}

\textbf{Network QoS Impacting Application QoE:} Prior works have studied how degraded network conditions have impacted the user experience of multimedia applications such as online gaming \cite{CSGO, networkedGaming, fps}, cloud gaming \cite{LyuIFIPNetworking, 10178055, lyu2024network, jarschel2011evaluation, laghari2019quality}, and video streaming \cite{9072143, barman2019qoe, nightingale20185g, nam2014towards, madanapalli2021reclive}. For example, the work in \cite{CSGO} measured the impact of latency on the first-person shooting game Counterstrike by analysing player experience at different latency levels. The authors rated their user experience using Mean Opinion Score (MOS) for each considered tested latency level. The authors of \cite{10178055} analysed how cloud gaming user experience (\eg freeze, frame rate and decoding delay) can be detected using machine-learning models. The study in \cite{9072143} analysed how bandwidth constraints impact start-up delays, resolutions and stalling occurrences in cloud games. The work in \cite{10148093} investigated the impact of bandwidth on video streaming performance.
We have noticed that no prior work has studied the impact of degraded network conditions for metaverse VR applications, which becomes the focus of our study.

\section{Conclusion}
Our work empirically evaluates the user QoE within metaverse VR public and private events impacted by degraded network QoS. We first systematically formalise three QoE metrics, including Freeze, Content Loading and Control Response Time that are assessed by five standard ACR ratings (\ie Excellent, Good, Fair, Poor, Bad). Focusing on both public social hubs and private user-generated events across three popular metaverse applications (Rec Room, VRChat, and MultiverseVR) on the Oculus VR Platform, we measure user experience through various conditions of three network QoS parameters (bandwidth, latency and packet loss) and benchmark minimum QoS values for each level of overall user QoE. Our study provides a reference for telecommunication network operators to prepare their networks to better support metaverse applications.

\bibliographystyle{IEEEtran}
\bibliography{reference}

\end{document}